\documentclass{ws-ijmpa}

\begin{document}

\markboth{Morgan O. Wascko}
{Measuring Charged-Current $\nu_{\mu}$ Interactions in MiniBooNE}

%
\catchline{}{}{}{}{}
%

\title{Measuring  $\nu_{\mu}$ Charged-Current Interactions in MiniBooNE}

\author{\footnotesize Morgan O. Wascko for the BooNE
  Collaboration\footnote{ email: \texttt{wascko@fnal.gov}}}

\address{Department of Physics and Astronomy, Louisiana State University,\\
Baton Rouge, LA, 70803,
U.S.A.
}

\maketitle


\begin{abstract}
  MiniBooNE seeks to confirm or refute the LSND
  $\overline{\nu}_{\mu}\rightarrow\overline{\nu}_e$ oscillation signal
  with high statistical significance and different systematics.
  MiniBooNE has accumulated the world's largest $\sim$1~GeV neutrino
  data set. MiniBooNE employs a cosmic muon calibration system to
  study the reconstruction of the energies and directions of muons in
  the detector.  Progress of measurements of the $\nu_{\mu}$
  charged-current quasi-elastic and single pion production cross
  sections are presented.
\end{abstract}

\section{The Importance of Measuring Muons in MiniBooNE}

MiniBooNE~\cite{boone-prop} is a neutrino oscillation experiment at
Fermilab designed to confirm or rule out the hypothesis that the LSND
$\overline{\nu}_e$ excess~\cite{lsnd} is due to $\overline{\nu}_{\mu}
\ \rightarrow \ \overline{\nu}_e$ oscillations. A general description
of the experiment can be found elsewhere in these proceedings~\cite{hray}.

The neutrino energy reconstruction is critical to the success of the
MiniBooNE oscillation and cross section analyses.  Charged current
quasi-elastic (CCQE) events ($\nu_{\mu}n\rightarrow\mu^- p$) are
typically used to measure the neutrino energy spectrum because they
have simple kinematics.  Neglecting the nucleon target momentum, the
reconstructed quasi-elastic neutrino energy can be expressed in terms
of the momentum of the muon.  Therefore, the muon energy and direction
measurements completely determine the neutrino energy measurement.

The MiniBooNE cosmic muon calibration system~\cite{calib-NIM} uses
stopping muons and their decay electrons to calibrate the event
reconstruction algorithms.  This system provides a precise calibration
of the energy, direction and position of muons for the complete range
of muon energies of interest in the experiment, 100-900~MeV.

\section{Cosmic Muon Calibration System}

The muon calibration system consists of a muon tracker located above
the detector, and seven scintillator cubes deployed inside the
detector.  The entering position and direction of a cosmic muon
impinging on the detector are determined by the muon tracker, and the
stopping position is determined by the location of the cube in the
case where the muon stops and decays inside the one of the cubes.  The
muon energy is then obtained from the range with an uncertainty due to
range straggling of approximately 3\%~\cite{stern}.  The muon range
kinetic energy measurement is compared to the visible energy as
reconstructed by the event fitters on an event by event basis.  This
gives the absolute energy scale calibration of the MiniBooNE detector.

\begin{figure}
    \hspace{-0.2in}
  \centerline{\psfig{file=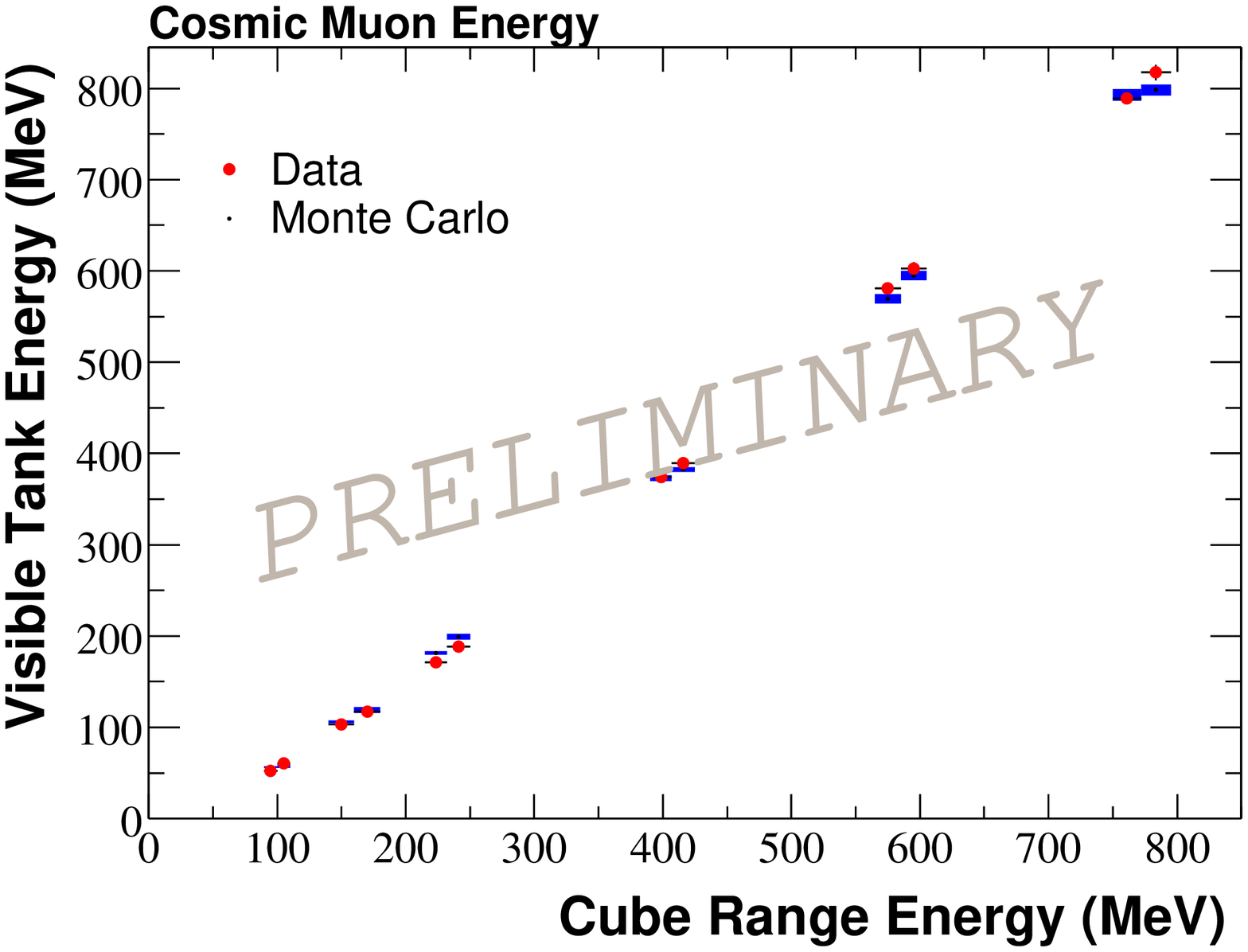,height=2.0in,width=2.3in}
    \hspace{0.2in}
    \psfig{file=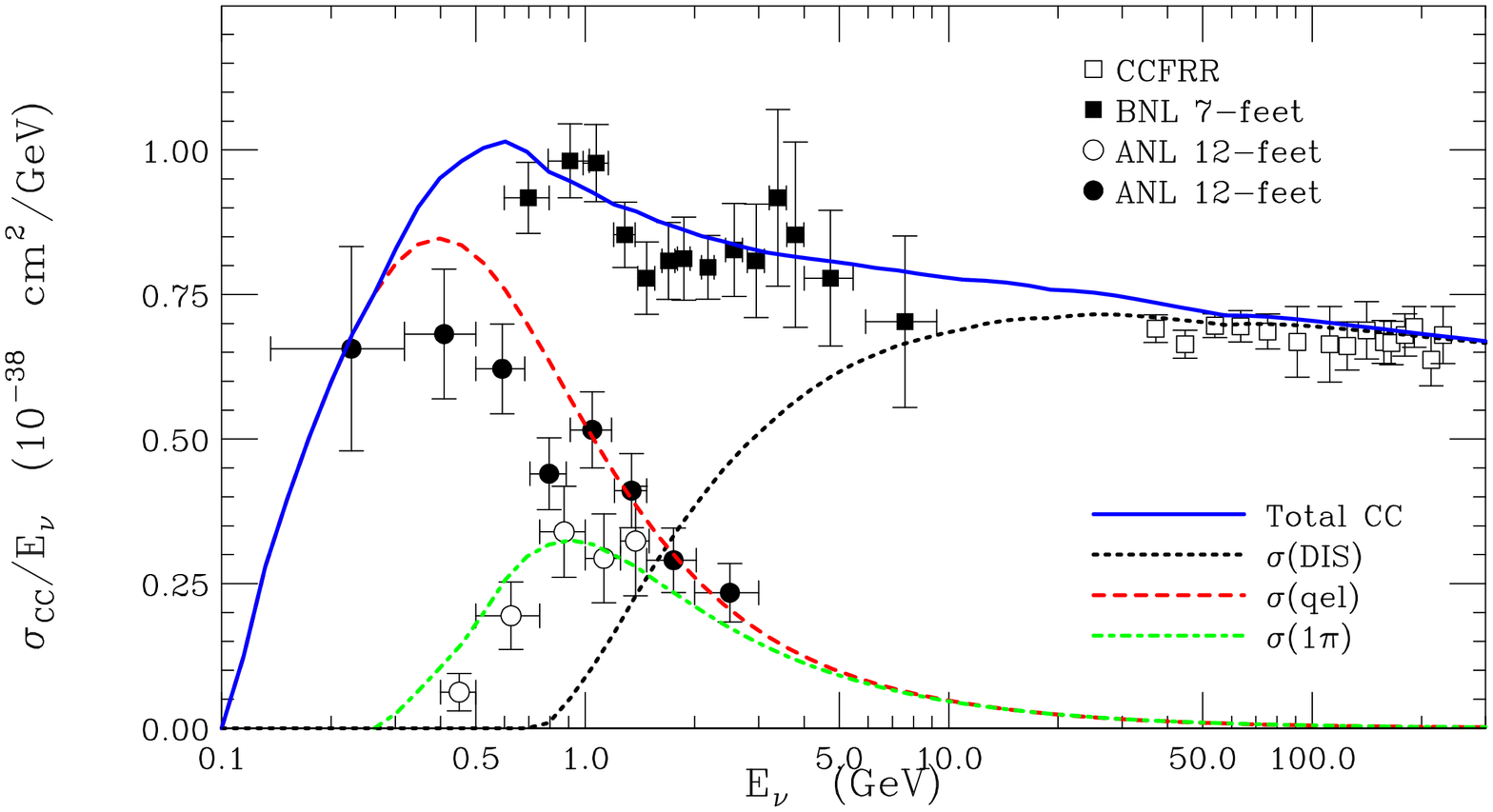,height=2.3in,width=2.0in,angle=90}}
\caption{\label{fig:cubes-xsec}The energy of cosmic muons in MiniBooNE, 
  and the $\nu_{\mu}$ interaction cross sections shown as a function
  of neutrino energy.}
\end{figure}

The event fitter returns an ``electron equivalent energy,'' which is
the energy of an electron that would have produced the same number of
photoelectrons in the detector~\cite{pdg1}.  The visible energy of
cosmic muons is plotted against the kinetic energy calculated from the
range using the cubes in Fig.~\ref{fig:cubes-xsec}a.  There is good
agreement between data and Monte Carlo.  Using the information in
Fig.~\ref{fig:cubes-xsec}a, the visible energy measurement is
converted to a muon kinetic energy, which is used to calculate the
neutrino quasi-elastic energy.  From the cosmic muon calibration
system, the muon kinetic energy uncertainty is measured to be 5\%, and
the angular resolution to be 45~mrad, leading to a neutrino energy
uncertainty of $\sim$10\%.

\section{Measuring $\nu_{\mu}$ Events in MiniBooNE}

The MiniBooNE event rate prediction is described
elsewhere~\cite{monroe-ccqe}.  The $\nu_{\mu}$ charged-current
interaction cross section as a function of neutrino energy is shown in
Fig.~\ref{fig:cubes-xsec}b~\cite{lipari}.  In MiniBooNE's energy
range, the dominant cross section are CCQE, which comprises 40\% of
the expected neutrino events, and charged current resonant single pion
(CC1$\pi$) events, which are expected to comprise 25\% of the total
event rate.

\subsection{Charged-Current Quasi-Elastic Events}

CCQE events are interesting because they are the dominant event
channel used in successful neutrino oscillation searches.  MiniBooNE's
CCQE event selection requires that candidate events pass cosmic
background and fiducial volume cuts, and that the event topology be
consistent with expectations for a single muon passing through the
detector~\cite{monroe-ccqe}.  Monte Carlo studies indicate the cuts
are 55\% efficient within the 500cm fiducial volume, producing an 80\%
pure CCQE event sample.

\begin{figure}
  \centerline{\psfig{file=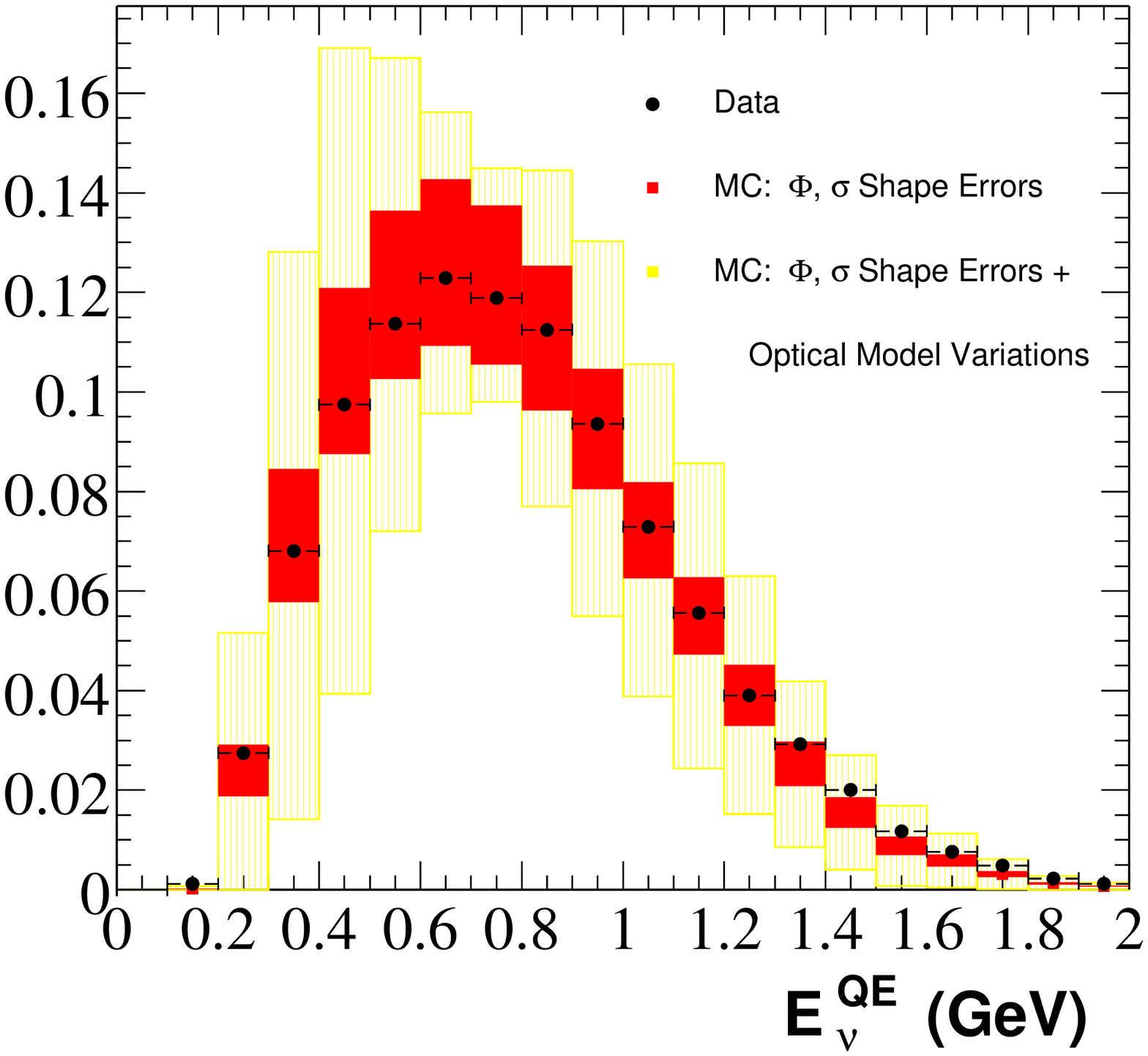,width=2.5in}
    \psfig{file=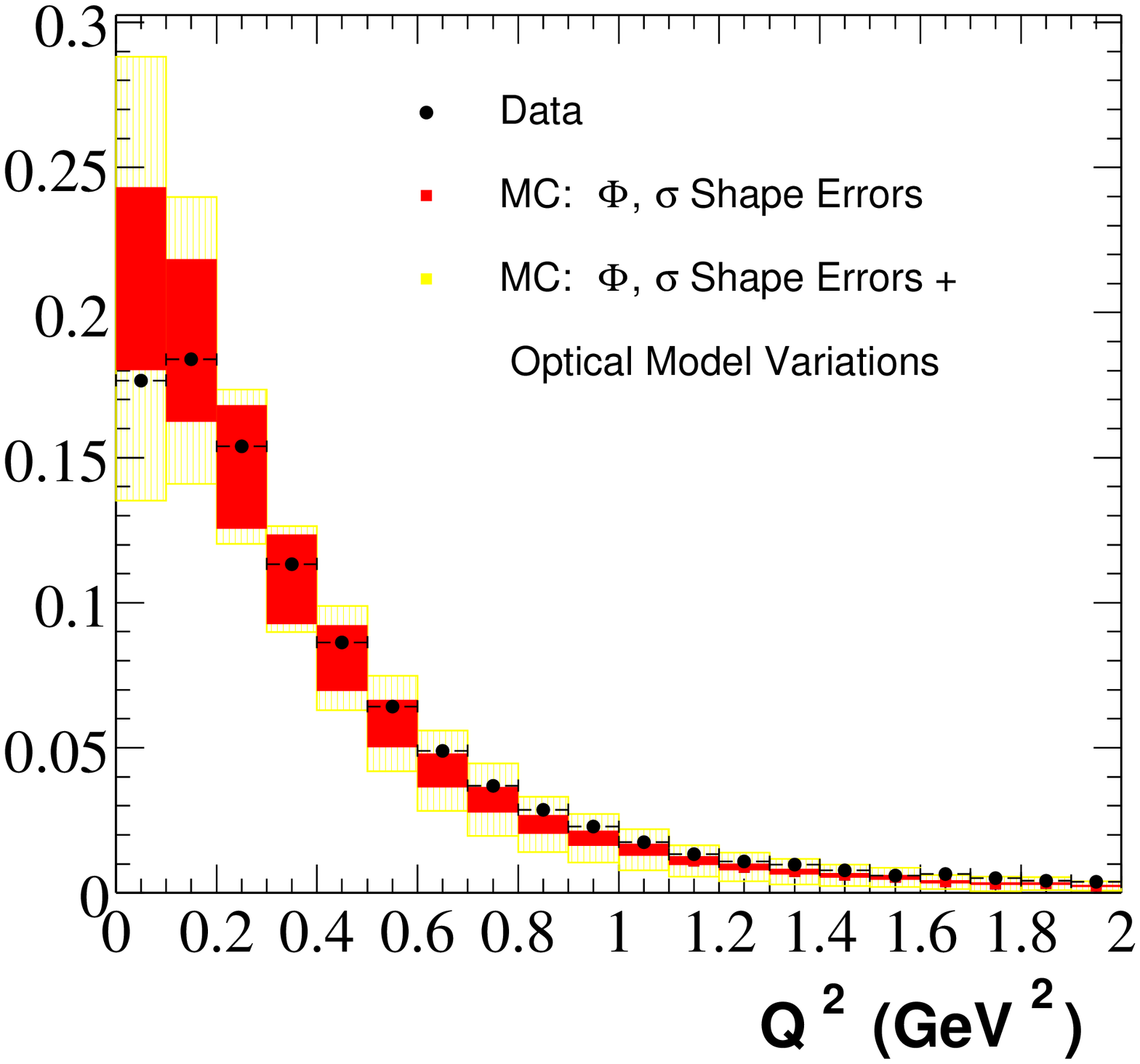,width=2.5in}}
\caption{\label{fig:ccqe}Distribution of reconstructed quasi-elastic energy of 
  MiniBooNE $\nu_{\mu}$ event candidates.  Distribution of
  reconstructed momentum transfer of MiniBooNE $\nu_{\mu}$ event
  candidates. The figures both show data in black points, with
  statistical errors, and Monte Carlo expectations in colored bands
  with systematic uncertainties.  }
\end{figure}

The reconstructed neutrino quasi-elastic energy is shown in
Fig.~\ref{fig:ccqe}a, and the Q$^2$ distribution in
Fig.~\ref{fig:ccqe}b, for 1.6$\times$10$^{20}$~POT.  Ongoing studies of
the transmission of light in mineral oil are expected to improve the
optical variations dramatically.

Note that the shape of the neutrino energy spectrum is somewhat harder
in the data than the Monte Carlo predicts, although the deviations sit
within the limits of the current error bands.  Note also the hint of a
low Q$^2$ deficit which may indicate a nuclear model deficiency.  This
is an active area of study within the neutrino community.

\subsection{Charged-Current Single Pion Events}

Charged-current single pion ($\nu_{\mu}p\rightarrow\mu^-\pi^+p$)
production has been studied since the advent of high energy
accelerator neutrino beams but the cross section for such processes in
the MiniBooNE energy range have not been sufficiently explored.  We
describe here the first look at a sample of CC1$\pi$ events in
MiniBooNE.

MiniBooNE's CC1$\pi$ event selection requires the simple yet robust
cut of two Michel electrons following the neutrino interaction.  The
majority of pions emitted from these events stop in the detector oil.
These decay quickly to muons, which then decay to Michel electrons.
The muons emitted from the neutrino interaction also come to rest, and
the majority of these decay to Michel electrons.  Applying this
requirement to 2.7$\times$10$^{20}$ POT of MiniBooNE data yields over
36,000 CC1$\pi$ candidate events.  The Monte Carlo predictions
indicate a purity of 80\% for this sample.  This data set is larger by
a factor of 4 than all CC1$\pi$ neutrino data published to date.


The Michel electrons from the CC1$\pi$ candidate events are used to
verify the composition of the data set.  
The distance from each Michel to the end of the
muon track in calculated.
Assuming that the closer Michel is associated with the $\mu^-$ from
the neutrino interaction, and the farther Michel is associated with
the $\pi^+$ decay, we expect the closer Michels to have a shorter
lifetime.  This occurs because the $\mu^-$ are captured by carbon
nuclei at a rate of 8\%, changing the expected lifetime from
2197.03$\pm$0.04~ns~\cite{mu-} to 2026.3$\pm$1.5~ns~\cite{mu+}.  The
observed muon lifetimes for the close and far Michel samples are
2070$\pm$16~ns and 2242$\pm$17~ns, respectively.  Again, note that the
observed lifetimes do not yet include systematic uncertainties.

While we are able to successfully extract CC1$\pi$ events with high
purity, full event reconstruction studies are still in progress as the
complex final state requires additional reconstruction handles that
are not yet fully developed.

\section{Conclusions}

MiniBooNE has already amassed the world's largest neutrino data set in
the $\sim$1~GeV region in its quest to confirm or rule out the LSND
oscillation signal.  Using a cosmic muon calibration system, we
measure the energy of muons to 5\%, and the directions to better than
45~mrad.  This leads to an uncertainty in the reconstruction of
quasi-elastic neutrino energy of $\sim$10\%.  We are currently
examining large CCQE and CC1$\pi$ data sets, and expect to have cross
section measurements and $\nu_{\mu}$ disappearance oscillation results
from these data samples in 2005.

\end{document}